\documentclass{llncs}
 
\usepackage{graphicx} 
\usepackage{dcolumn} 
\usepackage{bm} 
\usepackage{amsxtra} 
\usepackage{latexsym} 
\usepackage{comment}

\newenvironment{keywords}{
       \list{}{\advance\topsep by0.35cm\relax\small
       \leftmargin=1cm
       \labelwidth=0.35cm
       \listparindent=0.35cm
       \itemindent\listparindent
       \rightmargin\leftmargin}\item[\hskip\labelsep
                                     \bfseries Keywords:]}
     {\endlist}

\begin{document}

\title{Stochastic spreading processes on a network model based on regular graphs} 
 
\author{Sebastian V. Fallert\inst{1} \and Sergei N. Taraskin\inst{1}$^{,}$\inst{2}$^{,*}$} 
\institute{Department of Chemistry, University of Cambridge, 
  Cambridge, UK
  \and St. Catharine's College, University of Cambridge, Cambridge,
  UK, snt1000@cam.ac.uk\\\vspace{0.25cm}(* corresponding author)} 
 
\maketitle 
\begin{abstract} 
The dynamic behaviour of stochastic spreading processes on 
a network model based on k-regular graphs is investigated. 
The contact process and the susceptible-infected-susceptible model for 
the spread of epidemics are considered as prototype stochastic spreading 
processes. 
We study these on a network consisting of a mixture of $2$- 
and $3$-fold coordinated randomly-connected nodes of concentration $p$ 
and $1-p$, respectively, with p varying between 0 and 1. 
Varying the parameter  $p$  from  $p=0$  ($3$-regular graph of infinite 
dimension) to  $p=1$ ($2$-regular graph - 1D chain) allows us to 
investigate their  behaviour  under such structural changes. 
Both processes are expected to exhibit mean-field features for  $p=0$ 
and features typical of the directed percolation universality class for $p=1$ . 
The analysis is undertaken by means of  Monte Carlo simulations and 
the application of mean-field theory. 
The quasi-stationary simulation method is used to obtain the phase 
diagram for the processes in this environment along with critical exponents. 
Predictions for critical exponents obtained from mean-field theory are 
found to agree with simulation results over a large range of values 
for  $p$  up to a value of $p=0.95$, where the system is found to 
sharply cross over to the one-dimensional case. 
Estimates of critical thresholds given by mean-field theory are found 
to underestimate the corresponding critical rates obtained numerically 
for all values of $p$.  
\end{abstract} 


\section{Introduction}

The spread of epidemics poses a threat to biological populations as well as to 
computer networks and investigations into its dynamics and mechanisms are 
therefore of great current interest.  
One common class of epidemic models considers individuals to be in one 
of two 
possible states: susceptible (S) or infected (I). 
In this paper, we consider both the Contact Process (CP) 
\cite{harris_74} and the SIS 
model, two models of disease propagation via nearest neighbour 
contact, in which a disease is passed on to healthy nearest 
neighbours stochastically at a rate $\lambda$ specific to the model 
 while 
infected sites spontaneously recover at rate $\epsilon$. 
 
These Markovian spreading processes have 
attracted wide attention in the past due to their applicability to phenomena as  
diverse as autocatalytic chemical reactions, spreading of rumours and 
transport in disordered media \cite{marro_99:book}. 
As the rates $\lambda$ and $\epsilon$ are varied, an epidemic will 
be in one of two distinct states: an invasive regime (active 
state) in which it is present with a non-zero probability of 
ultimate survival and one in which this probability is zero thus 
leading to a state which allows no further evolution  
because the disease has died out (absorbing state).  
 
These two regimes are known to be connected by a continuous phase  
transition thereby rendering them of  
conceptual interest for investigations into this kind of  
critical phenomenon of non-equilibrium statistical mechanics  
(see \cite{Hinrichsen_00:review} for a review).  
The critical behaviour for these models in one-, two- and 
three-dimensional lattices has been investigated very 
accurately \cite{grassberger_89} and is found to be characteristic of 
the Directed Percolation (DP) universality class. 
From a range of studies, critical thresholds for the phase transition 
as well as critical exponents of predicted power-law scaling relations 
are known to high precision.  
 
With the growing interest in complex networks among the statistical physics 
community in recent years \cite{dorogovtsev_02,albert_02}, the 
question of the behaviour of dynamic processes 
on such topologically disordered structures has arisen \cite{dorogovtsev_03,dorogovtsev_07}. Particularly motivated 
by the fact that networked structures are ubiquitous in nature, the 
effects of these environments on, for example, the spread of a disease are 
of immediate interest. In a series of papers 
\cite{Pastor-Satorras2001,Pastor-Satorras2001a,Boguna2003,Castellano2006,Park2007}, 
the behaviour of the CP and the SIS model on a range of networks has been 
considered and even comparisons with data of computer virus outbreaks have 
been attempted \cite{Boguna2003}. As networks in general are infinite-dimensional 
objects, the dynamical mean-field (MF) approximation is expected to become exact in 
these cases in principle rendering many models tractable by analytical 
means. Both Monte Carlo (MC)  
simulations and the MF approximation have been used in previous investigations 
and produced such astonishing results as the absence of an epidemic threshold 
infection rate for infinite scale-free networks \cite{Boguna2003}.  
 
In this paper, 
 we propose to investigate the behaviour of the CP and 
the SIS model as two paradigmatic stochastic spreading processes on 
networks of  k-regular graph topology. 
The model network considered in this investigation consists of a 
mixture of $2$- and $3$-fold coordinated randomly-connected nodes of  
concentration $p$ and $1-p$, respectively.
Varying the parameter  $p$  from  $p=0$  to $p=1$ transforms the 
system from a $3$-regular graph of infinite dimension to a $2$-regular 
graph, i.e. a 1D chain.  
While both the CP and the SIS model are expected to exhibit mean-field  
features for  $p=0$, the processes effectively take place in a 
one-dimensional environment for $p=1$ which is a very well-studied
regime of the DP universality class. 
It is our aim to investigate the behaviour of both the critical rates 
and accessible critical exponents 
for this crossover from an 
infinite- to a one-dimensional case thereby probing the validity of 
the MF approximation in this setting.  
The analysis is undertaken by means of Monte Carlo simulations using 
the quasi-stationary (QS) simulation method \cite{de_oliveira_05} and 
the application of mean-field theory \cite{Pastor-Satorras2001a}. 
 
This paper is structured as follows. Section \ref{sec:bg} outlines the 
definitions and some properties of the processes considered. The MF 
approximation and the QS simulation method are described in section 
\ref{sec:methods}. We present and discuss our results in section 
section \ref{sec:results} and summarise our findings in section 
\ref{sec:conclusion}.

\section{\label{sec:bg}Background} 
 
As outlined in the previous section, both the CP and the SIS model are 
simple toy models for the spread of an infectious disease by nearest-neighbour 
contact.  
In these models 
defined on a network, nodes represent susceptible or infected
 individuals 
surrounded by their neighbours connected via links along which the 
epidemic may spread. 
Proliferation of the disease to nearest neighbour sites happens at a 
transmission rate $\lambda$ while recovery is spontaneous at rate 
$\varepsilon$ making the sequence of events an individual can cycle 
through $\textrm{S}usceptible \rightarrow \textrm{I}nfected \rightarrow\textrm{S}usceptible$. 
 
The CP and the SIS model are very similar, the difference being the 
exact mechanism of the spreading of infection. 
In the case of the CP, a site attempts to transmit its disease at rate 
$\lambda/k$ to a randomly selected neighbour where $k$ denotes the number 
of nearest neighbours. 
If the selected neighbour is already infected, 
proliferation fails.
For the SIS model, transmission to any non-infected 
neighbour happens,  in contrast, at rate $\lambda$ independent 
of the connectivity of  the nodes. 
Thus, the spreading mechanism in the CP effectively compensates for 
the local connectivity present in the network through a suitable 
reduction of the spreading rate through a particular link.  
 
 
Once suitable initial states for all sites have been 
chosen, the above rules dynamically evolve the spread of a disease in  
the network.  
A typical initial condition is the state of a fully-infected system 
from which the system relaxes very quickly.  
For very long times, and formally as 
time $t \to \infty$ and for an infinite number of sites in the network 
$N \to \infty$,  
the system is expected to be in one of two 
states: An \emph{active state} in which there remains a finite density 
of infected sites 
or an \emph{absorbing state} in which the disease has died out and 
that therefore 
admits no further time evolution. Depending on the value of the 
transmission and recovery rates $\lambda$ and $\epsilon$, ultimately the 
system will be in one of the two possible states.  
More precisely, there exists a continuous phase transition between these 
regimes as one fixes one of the rates and varies 
the other.  
This transition takes the system from a phase where the density of 
infected sites (order parameter) $\rho$ is zero to one where it 
continuously grows from zero as the transmission rate (control  
parameter, assuming $\varepsilon$ fixed) is 
increased. 
  
Without loss of generality, one can perform a rescaling of time and 
set one of the two rates to unity for convenience.  
In the following, the recovery rate is assumed to be $\epsilon=1$ 
and the critical point is therefore characterised by a critical 
transmission rate $\lambda_c$ alone. 
 
 
There exist a range of well-established scaling relations for various 
observables in these models of which we present those relevant for 
this investigation. 
The density of infected sites in the thermodynamic limit as $t \to 
\infty$, the order parameter, is expected to scale as 
\begin{eqnarray} 
\lim_{t \to \infty} \langle \rho(t) \rangle = \overline{\rho} \sim |\lambda-\lambda_c|^{\beta} 
\end{eqnarray} 
thereby defining the order parameter critical exponent $\beta$ where 
$\langle \dots \rangle$ denotes averaging over realisations of the process. 
Order parameter fluctuations are known to follow 
\begin{eqnarray} 
V = N \left( \overline{\rho^2} - \overline{\rho}^2 \right) \sim |\lambda-\lambda_c|^{-\gamma} 
\end{eqnarray} 
%
%

%
%
%
%
Both the models under consideration are known to belong to the 
directed percolation (DP) universality class \cite{marro_99:book}. 
Accordingly, the critical exponents defined above are those 
characteristic of this universality class. 
Above the upper critical dimension, $d_u=4$, of these models, 
fluctuations are expected to be Gaussian and MF theory should be exact. 
Therefore, these processes taking place in infinite-dimensional 
networks are expected to exhibit exponents predicted by mean-field 
theory.

\section{\label{sec:methods}Methods} 
 
\subsection{Mean-Field Approximation}

Both the CP and the SIS model can be described by the master equation  
which  reflects 
the conservation of probability flow 
\cite{Hinrichsen_00:review}. 
In the following, we will first outline the case of the SIS model and 
then consider the extension to the simpler case of the CP. 
 
In the dynamical MF approximation, which neglects density fluctuations 
and statistical correlations between the densities at different sites, 
and, for the moment, disregarding the structure of the network 
completely, the master equation for the SIS model takes the form 
\begin{eqnarray} 
\partial_t \rho(t) = -\rho(t) ~ + ~ 
\lambda ~k~(1-\rho(t))~\rho(t) 
\label{eq:master_equation} 
\end{eqnarray} 
where $\rho(t)$ denotes the density of 
infected sites
at time $t$ averaged over 
realisations of the process which is identical to the probability of  
a site of the system to be infected at time $t$. 
This equation describes the 
rate of change of the 
average density in the network which is equal to the flow of density 
into and out of any site with time and makes for the destruction and the 
creation terms above.  
The destruction term due to the vanishing of infection at 
unit rate is proportional to the density $\rho(t)$.  
The creation term is due to the possible infection by infected 
neighbouring sites in the case that the vertex under consideration is 
not infected.  
Accordingly, it is proportional to the probability that a site is not 
infected, $(1-\rho(t))$, the probability that a neighbouring site 
is infected $\rho(t)$, the local connectivity $k$ and the spreading 
rate $\lambda$.

The master equation (\ref{eq:master_equation}) can be extended in 
order to take into account the structure of the underlying 
network at the level of the 
node degree (connectivity) distribution (as developed by 
Pastor-Satorras and Vespignani \cite{Pastor-Satorras2001}).  
It is clear that, unless 
one assumes a homogeneous network with $\langle k \rangle \approx k$ for all $k$, the 
expression will decouple into a set of equations 
for the densities of infected vertices characterised by a certain
connectivity $k$, 
We can write  for each $k$,  
\begin{eqnarray} 
\partial_t \rho_{k} = -\rho_{k} ~ + ~ 
\lambda ~k~(1-\rho_{k}(t))~\Theta_{k}(t)~, 
\label{eq:master_equation_k} 
\end{eqnarray} 
where $\Theta_{k}(t)$ is the probability that an edge emanating from a vertex 
of degree $k$ is connected to an infected site.  
The infection term as described above now incorporates the probability 
that a site of degree $k$ is connected to an infected vertex 
$\Theta_{k}(t)$.  
One can interpret $\Theta_{k}(t)$ as 
the mean density of neighbouring infected nodes and consequently $k 
\Theta_{k}(t)$ as the mean number of infected nearest neighbours  
\cite{Pastor-Satorras2001}.  
 
Networks which are Markovian are statistically described by their degree 
distribution $P\,(k)$ and the conditional probability 
$P\,(k'|k)$ that an edge of a node of degree $k$ is connected to a vertex of 
degree $k'$. 
For such systems, we can write 
\begin{eqnarray} 
\Theta_{k}(t) = \sum_{k'}~ P\,(k'|k) ~\rho_{k'}\,(t) 
\label{eq:theta} 
\end{eqnarray} 
where the sum runs over all degrees $k'$.  
For uncorrelated networks, which we will exclusively 
consider in this 
investigation, this becomes \cite{dorogovtsev_03} 
\begin{eqnarray} 
\Theta(t) = \sum_{k'}~ \frac{k'P\,(k')~ \rho_{k'}\,(t)}{\langle k \rangle} 
\label{eq:theta_nc} 
\end{eqnarray} 
which does not depend on $k$ any longer.  
Substituting this expression into the rate equation 
(\ref{eq:master_equation_k}) and imposing stationarity ($\partial_t \rho_k=0$) 
one obtains 
\begin{eqnarray} 
\overline{\rho}_k = \frac{k\lambda\overline{\Theta}}{1+k\lambda\overline{\Theta}} 
\label{eq:stat_rho} 
\end{eqnarray} 
where $\overline{\rho}_k$ and $\overline{\Theta}(\lambda)$ 
are the time-independent 
values for the mean density at sites of degree $k$ and the mean density of 
infected neighbouring sites.  
Multiplying by $\frac{P(k)k}{\langle k \rangle}$ and summing over 
$k$ yields 
\begin{eqnarray} 
\frac{1}{\lambda} = \frac{1}{\langle k \rangle} \sum_{k} 
\frac{P(k)~k^2}{1+k\lambda\overline{\Theta}} \equiv f(\lambda\overline{\Theta}) 
\label{eq:lambda_crit_1} 
\end{eqnarray} 
where $f(x)$ is a 
monotonically 
 decreasing function of $x$.  
This equation only has 
a (unique) solution for $\overline{\Theta}(\lambda)$ different from zero  
for $\lambda>\lambda_c$ where $\lambda_c$ is the threshold value for the  
transmission rate that makes $\overline{\Theta}$ smallest.  
As by definition the mean density of infected 
nearest neighbours $\overline{\Theta}\geq 0$ in the active regime and
$f(x)$ is a monotonically  
decreasing function we have (effectively setting $\overline{\Theta}=0$
\begin{eqnarray} 
\lambda_c^{\text{SIS}} = \frac{1}{f(0)} = \frac{\langle k \rangle}{\langle k^2 \rangle} 
\label{eq:sis_threshold} 
\end{eqnarray} 
for the critical threshold \cite{joo_04}.

In order to obtain an expression for the order paramter we start by
combining equations (\ref{eq:theta_nc}) and (\ref{eq:stat_rho}) and
arrive at a self-consistency equation for $\overline{\Theta}$ 
(equivalent to equation (\ref{eq:lambda_crit_1}))  
\begin{eqnarray} 
\overline{\Theta} = \sum_{k}~ \frac{kP(k)} {\langle k \rangle} 
~ \frac{k\lambda\overline{\Theta}}{1+k\lambda\overline{\Theta}} 
\label{eq:self_consistency} 
\end{eqnarray} 
that can in principle be solved for $\overline{\Theta}$ which in turn allows 
one to obtain the order parameter in the MF approximation from 
\begin{eqnarray} 
\overline{\rho} = \sum_{k}~ \overline{\rho_k}~.
\label{eq:rho_mf} 
\end{eqnarray} 
 
Critical exponents can be extracted from MF theory by considering the 
leading behaviour of the relevant expressions. 
For example, combining the last expression 
Eq.~(\ref{eq:rho_mf}) and Eq.~(\ref{eq:stat_rho}) and expanding in
$\lambda-\frac{\langle k \rangle}{\langle k^2 \rangle}$ in analogy to
the scaling form $\overline{\rho} \sim (\lambda-\lambda_c)^\beta$, one
obtains the MF value for the order parameter exponent $\beta=1$. 
Similarly, one obtains $\gamma=0$ for the corresponding fluctuations.
 
In the case of the CP where the effective spreading rate is inversely 
proportional to the number of links connected to an infected site, 
Eq.~(\ref{eq:theta}) has to be modified and reads 
\begin{eqnarray} 
\Theta_{k}(t) = \sum_{k'}~ \frac{P\,(k'|k) ~\rho_{k'}\,(t)}{k'} 
\label{eq:theta_CP} 
\end{eqnarray} 
which for uncorrelated networks leads to $\Theta^{nc}(t) = \rho(t) / 
\langle k \rangle$ where $\rho$ is the average density of 
infected sites  
averaged over degrees. 
Following the procedure as for the SIS model, the critical threshold 
rate is found to be degree independent and given by \cite{Castellano2006}
\begin{eqnarray} 
\lambda_c^{\text{CP}} = 1~
\label{eq:cp_threshold} 
\end{eqnarray} 
while the critical exponents are identical to those for the SIS model.
 
\subsection{Monte Carlo Simulation: Quasi-Stationary Simulation} 
 
Both processes under consideration can be simulated effectively  
via time-dependent Monte Carlo simulations. 
Once initial conditions have been chosen, the system is evolved 
according to the appropriate rules with a simulation that selects 
possible events according to their prescribed rates and ensures that 
they happen in exponentially-distributed time intervals. 
 
In contrast to lattices, networks are characterised by the existence 
of long range links. 
This implies that a dynamical process will very strongly feel the size 
of the system in a finite representation of the network leading to 
much stronger finite-size effects than experienced in lattices. 
While the most precise method of determining the critical point in 
lattices is spreading from a single seed while ensuring that the 
infection never reaches the boundary of the system, this is virtually 
impossible in networks. 
Thus, finite-size effects have to be systematically exploited in order 
to make a prediction about the infinite system using networks of 
different sizes. 
This is possible using the  finite-size scaling hypothesis which
predicts that values of observables in systems of size $L$ are 
controlled by the ratio $L/\xi_{\perp}$,  
where $\xi_{\perp}$ is the spatial correlation length. 
In order to use this scaling behaviour, one requires  
a stationary average value of these observables and analyses their
values for different system sizes.
Problematically, due to the  existence of the absorbing 
state no such true stationary state exists in a finite system. 
Fortunately, the processes under consideration evolve such that some 
observables attain quasi-stationary (QS) values.

Most notably, the density of 
infected sites averaged over surviving 
realisations of the process $\langle \rho(t) \rangle$ exhibits this 
behaviour after an initial transient starting from the initial state  
of a fully-infected system \cite{marro_99:book}. 
This quasi-stationary density $\overline{\rho}$ is expected 
to scale systematically in accordance with the finite-size scaling 
hypothesis. 
According to recent investigations into the finite-size scaling 
behaviour above the upper critical dimension \cite{luebeck_05,Park2007} the 
prediction is 
\begin{eqnarray} 
\overline{\rho} \sim L^{d\beta/2} ~ G\left(L^{d/2}(\lambda-\lambda_c) 
\right) = N^{\beta/2} ~ G\left (N^{1/2}(\lambda-\lambda_c) 
\right) 
\label{eq:park_scaling} 
\end{eqnarray} 
where $L^d=N$, the number of sites in the system and $G(x)$ an
appropriate scaling function.
A similar expression with $\beta$ replaced by $\gamma$ is valid for
the associated fluctuations.  
 
In principle the QS state can be investigated via a conventional 
simulation in which the system is stochastically evolved in time from 
a fully-infected initial condition. 
The density of infected sites conditioned on survival can then be 
analysed and a temporal average over the duration of the QS state is 
an estimator for $\overline{\rho}$. 
This method is however plagued by a range of problems \cite{Luebeck_2003} 
which led de Oliveira and Dickman to propose a simulation method which 
samples the QS state directly \cite{oliveira_2005}. 
 
In this QS simulation method, the absorbing state is eliminated and 
its probability 
weight is redistributed over the active states according to 
the history of the process. 
It can then be shown that the true stationary state of the resulting 
modified process corresponds to the quasi-stationary state of the 
original one. 
This method is ideally suited for our study as a single realisation of 
the network is investigated in one QS simulation run enabling us to 
analyse sample-to-sample fluctuations between realisations 
and find a critical point by use of the scaling relation Eq.~(\ref{eq:park_scaling}).

\section{\label{sec:results}Results and Discussion}

\subsection{MF Solution}

\begin{figure}[!ht]  
\centerline{\includegraphics[scale=0.35,angle=270]{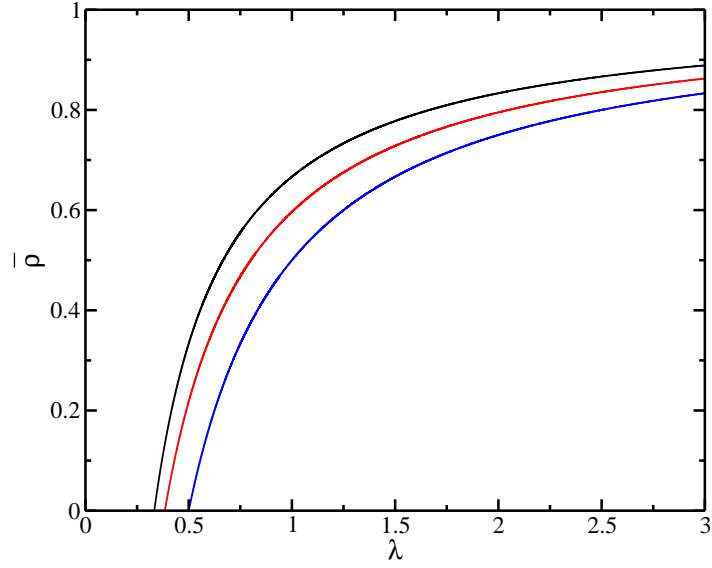}}  
\caption{The mean-field average density of infected sites for the SIS
  model on a binary 
  random network with $k_1=2$, $k_2=3$ for (from top to bottom) p=0, 0.5, 1.} 
\label{fig:mf_density}  
\end{figure}  
 
The mean-field theory for the CP and the SIS model on networks can be  
applied to our network 
with two types of nodes of connectivity $k_1$ and $k_2$ present with 
probabilities $P(k_1)$ and $P(k_2)$ respectively.  
For the SIS mode, the self-consistency equation, Eq.~\ref{eq:self_consistency}, for the  
stationary value of 
$\Theta$, the average density of infected nearest neighbours, takes the form 
\begin{eqnarray} 
\frac{\langle k \rangle}{\lambda} =  \frac{k_1^2 P(k_1)} 
{1+k_1 \lambda \overline{\Theta}} +  \frac{k_2^2 P(k_2)} {1+k_2 \lambda \overline{\Theta}} 
\label{eq:self_consistency_1} 
\end{eqnarray} 
which can be solved for $\overline{\Theta}$ giving 
\begin{eqnarray} 
\overline{\Theta} =  \frac{1}{2\lambda} \left( \lambda - \frac{k_1+k_2}{k_1 k_2} \right) 
+ \frac{1}{\lambda} \sqrt{ \frac{1}{4} ~  \left( \lambda - \frac{k_1+k_2}{k_1 k_2} 
\right)^2 + \frac{\langle k^2 \rangle \lambda - \langle k \rangle}{k_1 k_2 \langle k \rangle} } 
\label{eq:self_consistency_2} 
\end{eqnarray} 
where $\overline{\Theta}$ is only defined for transmission rates 
$\lambda>\lambda_c=\frac{\langle k \rangle}{\langle k^2 \rangle}$ as 
explained above.  
Using the definition of the average stationary 
density $\rho=\sum_k P(k) \rho_k$ and finally substituting $P(k_1)=p$ and 
$P(k_2)=1-p$, the order parameter in the MF approximation is given by 
\begin{eqnarray} 
\overline{\rho} = \lambda \overline{\Theta} \left( \frac{p k_1} 
{1+k_1 \lambda \overline{\Theta}} +  \frac{(1-p) k_2} {1+k_2 \lambda \overline{\Theta}} \right) 
\end{eqnarray} 
 
The critical exponents given by the leading order contributions of the 
relevant expressions are found to be the standard MF exponents
\cite{Castellano2006} as outlined above.
 
This solution is plotted for the special case $k_1=2$ and $k_2=3$ with $p$ 
varying from 0 to 1 in Fig.~\ref{fig:mf_density}. As expected, MF 
theory qualitatively reproduces the features of the phase transition: There 
exists a critical rate $\lambda_c$ below which the stationary density is zero while it 
grows continuously for values above.  
For the CP, an analogous analysis yields a similar solution.

No direct comparison with numerical predictions for the order 
parameter is shown in Fig.~\ref{fig:mf_density} because of severe
finite-size effects for networks which shift the simulation curve well
above its true asymptotic position. 
We will however compare the critical threshold rate as well as some 
critical exponents in the next section. 
 
\subsection{Simulation Results} 
 
\begin{figure}[!ht]  
\centerline{\includegraphics[scale=0.35,angle=270]{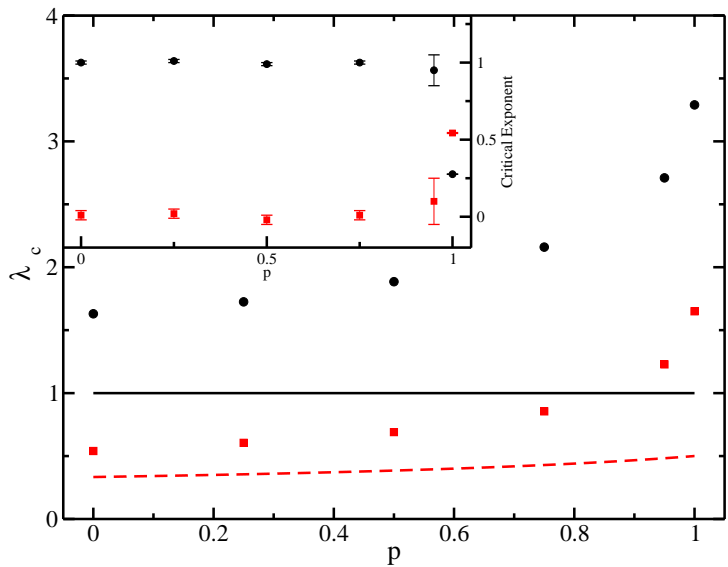}}  
\caption{The critical threshold $\lambda_c$ obtained from MF theory 
  (lines) and MC simulation (circles and squares) for the CP (circles and 
  solid line) and the SIS model (squares and dashed line) for various
  values of $p$. Inset shows the
  critical exponents $\beta$ (circles) and $\gamma$ (squares) as
  a function of $p$.} 
\label{fig:thresholds}  
\end{figure}  
 
The critical thresholds $\lambda_c$ for a range of values of $p$ for 
both the CP and the SIS model were obtained via the QS simulation method 
outlined above.  
We used networks of sizes ranging from $N=256 - 32768$ 
in QS simulation runs up to $10^8$ time steps averaging over no less 
than $100$ and up to a maximum of $1000$ network realisations (the
latter were required to minimise errors in light of strong 
sample-to-sample fluctuations for large values of $p$).  
The resulting critical thresholds are shown in 
Fig.~\ref{fig:thresholds} along with the MF predictions obtained from 
Eqs.~(\ref{eq:sis_threshold}) and (\ref{eq:cp_threshold}). 
As expected from the definition of the two models, the CP threshold 
exceeds the one of the SIS model for a particular value of $p$ which can be 
attributed to the reduction of the effective transmission rate by the  
local coordination number as in Eq.~(\ref{eq:theta_CP}). 
Also, in the two cases of homogeneous connectivity, $p=0$ and $p=1$, the 
thresholds for the two models are expected to be simply related by a 
factor of $3$ and $2$ (equal to the connectivity), respectively, 
 as can
be seen from Fig.~\ref{fig:thresholds}. 
Note that the critical threshold for the SIS model in the 
quasi one-dimensional case ($p=1$), $\lambda_c=1.65$, is almost identical to the 
threshold for the CP on the 3-regular random network ($p=0$) 
$\lambda_c=1.63$. 
 
Turning to the MF predictions in comparison to the MC results, one 
notes that they underestimate the true critical thresholds for 
all values of $p$ and for both models. 
The difference between the MF approximation and the simulation results 
is more pronounced for the CP as compared to the SIS model. 
\begin{figure}[!ht]  
\centerline{\includegraphics[scale=0.35,angle=270]{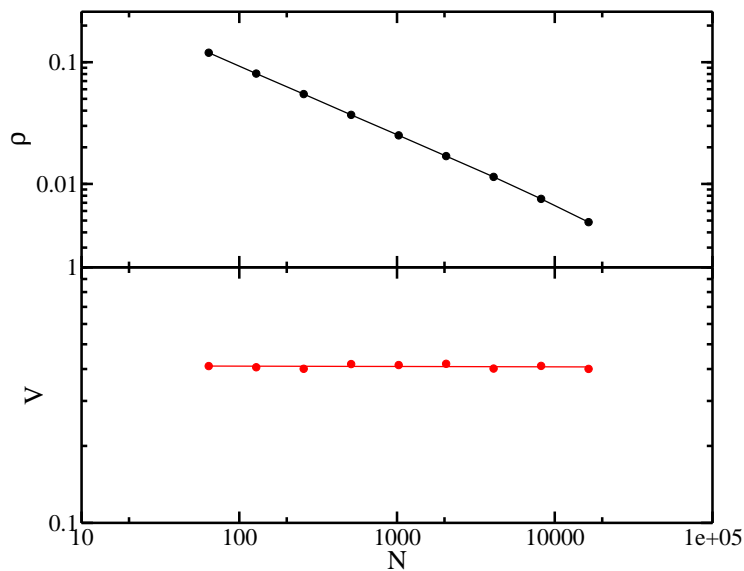}}  
\caption{The density of infection $\rho$ (upper panel) and the 
  corresponding fluctuations $V = N \left( \overline{\rho^2} - 
    \overline{\rho}^2 \right)$ (lower panel) for various network 
  sizes for the case $p=0.5$ in the SIS model. Solid lines are
  best-fit regression lines to the scaling forms defined in 
Eq.~(\ref{eq:park_scaling}).} 
\label{fig:typical}  
\end{figure}  
 
The exponents $\beta$ and $\gamma$ were obtained by fitting data for 
the density of infected sites $\rho$ and the corresponding fluctuations 
$V = N \left( \overline{\rho^2} - \overline{\rho}^2 \right)$ to the 
finite-size scaling form of Eq.~(\ref{eq:park_scaling}). 
A typical set of data points is shown in Fig.~\ref{fig:typical} for 
the case $p=0.5$ for the SIS model along with best-fit regression lines. 
Both quantities as a function of network size $N$ show power-law 
behaviour with the expected MF exponents $\beta/2=0.5$ and 
$\gamma=0$ indicating the validity of the MF approximation 
for this case.
These values of critical exponents are plotted in the inset of
Fig.~\ref{fig:thresholds}. 
As $p$ is further increased, strong sample-to-sample fluctuations 
for values beyond $p=0.95$ render a precise analysis very complicated.
For $p=1$, the well-established 1D finite-size scaling exponents are 
recovered ($\beta/\nu_{\perp}=0.253$ for $\rho(N)$ and $\gamma / 
\nu_{\perp}=0.498$ for $V(N)$ \cite{marro_99:book}) as can be seen
from 
the figure.
In the transition region, error bars for exponents are large and our
results give slight preference to the scenario of a discontinuous
change in exponent values.
However, we feel that a very rapid yet continuous change of exponents 
towards the 1D values cannot be excluded.

\section{\label{sec:conclusion}Conclusion} 
 
We have investigated the CP and the SIS model, two paradigmatic 
stochastic spreading processes, in a network model which 
interpolates smoothly between an infinite-dimensional 3-regular random network 
and a linear chain through the variation of a single parameter $p$. 
The MF approximation yields a prediction for the critical threshold 
rate of an epidemic outbreak and critical exponents associated with the 
corresponding absorbing state phase transition. 
For no value of $p$ does MF theory predict the true critical threshold 
as calculated from MC simulations. 
The predictions for critical exponents agree perfectly with 
simulations for a very wide range of $p$ up to $p=0.95$. 
Beyond this point, the analysis is complicated by strong
sample-to-sample fluctuations.
For $p=1$ one recovers the established exponents for the
one-dimensional case indicating a sudden crossover. 
While not being able to investigate the nature of this transition
precisly, our simulations favour the scenario of a discontinuous
change in the scaling exponents which reflects the abrupt change of the
dimensionality of the network.

\vspace{0.5cm} 
 
\emph{Acknowledgements} - MC simulations were performed 
on the Cambridge University Condor Grid. SVF acknowledges financial 
support from the EPSRC and the Cambridge European Trust. 
 
\bibliographystyle{splncs} 

\end{document}